\documentstyle[prl,amsmath,amssymb,amsthm,aps,epsf]{revtex}

\theoremstyle{remark}
\newtheorem{rem}{Remark}[section]

\begin{document}
\draft
\title{
\large \bf
Breaking of the overall permutation symmetry in nonlinear optical
susceptibilities of one-dimensional periodic dimerized
H\"{u}ckel model
}
\author{Minzhong Xu\thanks{Author to whom correspondence should be addressed.
Email: mx200@nyu.edu}}
\address{
Department of Chemistry, New York University, New York, NY 10003
        }
\author{Shidong Jiang}
\address{
Department of Mathematical Sciences, New Jersey Institute of Technology,
Newark, NJ 07102
}
\date{\today}
\maketitle
\bigskip
\begin{abstract}
Based on one-dimensional single-electron infinite periodic models of
trans-polyacetylene, we show analytically that the overall permutation
symmetry of nonlinear optical susceptibilities is, although preserved in 
bound-state molecular systems, no longer generally
held in periodic systems. The overall permutation
symmetry breakdown provides a natural explanation of the widely
observed large deviations of Kleinman symmetry in off-resonant regions
of periodic systems. Finally, physical conditions to experimentally test
the overall permutation symmetry breakdown are discussed.
\end{abstract}

{
The $n$th-order optical susceptibility is generally defined as a 
rank $n$ tensor
$\chi^{(n)}_{\mu\alpha_1\alpha_2\cdots\alpha_n}
(-\omega_\sigma;\omega_1,\omega_2,\cdots,\omega_n)$, 
where $\omega_\sigma\equiv\displaystyle\sum_{i=1}^n\omega_i$ is the sum of incoming 
frequencies and
$\mu\alpha_1\cdots\alpha_n$ are the indices of spatial directions. 
The intrinsic
permutation symmetry, as described in Butcher and Cotter's book\cite{butcher},
implies that the $n$th-order susceptibility is invariant under 
all $n!$ permutations
of pairs $(\alpha_1,\omega_1)$, $(\alpha_2,\omega_2)$,$\cdots$, 
$(\alpha_n,\omega_n)$. Intrinsic symmetry is a fundamental property of the
nonlinear susceptibilities which arises from the principles of time
invariance and causality, and applies universally to all physical systems. 
For the medium that
is transparent and lossless for all relevant frequencies, i.e., far away 
from all transition frequencies, it is generally believed that 
the optical susceptibilities have a much more 
interesting property, namely, the overall permutation symmetry
(or the full permutation symmetry in Boyd's book\cite{boyd}), in which the 
susceptibilities are invariant 
when the permutation includes the additional pair $(\mu,-\omega_\sigma)$. 
Therefore, the $n$th-order susceptibility is invariant under all $(n+1)!$
permutations of the pairs $(\mu,-\omega_\sigma)$, $(\alpha_1,\omega_1)$,
$\cdots$, $(\alpha_n,\omega_n)$\cite{butcher64}. 
Furthermore, when
the optical frequencies are much smaller than any of the transition 
frequencies, the dispersion of the medium at the relevant frequencies is
negligible. It follows that the susceptibility is asymptotically
invariant under all
permutations of the subscripts $\mu$, $\alpha_1$, $\cdots$, $\alpha_n$ in 
the low frequency limit. This property is known as Kleinman 
symmetry\cite{butcher,boyd,kleinman}. 

However, as observed by Simpson and his coauthors\cite{simpson04}, the 
overwhelming majority of recent optical experiments on organic
materials and crystals showed large deviations from Kleinman symmetry,
even in the low-frequency off-resonant regions. Although the deviations
from Kleinman symmetry in molecular systems are fairly small. In this
paper, based on the theoretical 
framework developed in our previous work\cite{mxu1,mxu2,mxu3,mxu4,jiang1}, we 
prove that the overall permutation symmetry of nonlinear optical 
susceptibilities is broken in one-dimensional (1D) periodic systems.
On the other hand, the overall permutation symmetry remains valid in
bound-state molecular systems.
Since the overall permutation symmetry is the basis of Kleinman symmetry, this 
provides a natural explanation why the deviations from Kleinman symmetry is 
much larger in periodic systems than in bound-state molecular
systems. Indeed, despite the wide acceptance of the overall 
permutation symmetry in 
the nonlinear optics\cite{butcher,boyd,butcher64}, no direct measurement has 
tested the validity of the assertion. We will suggest physical
conditions to experimentally test the overall permutation symmetry breakdown. 

The analytical derivations of the overall permutation symmetry of
nonlinear susceptibilities are rigorous 
and correct in molecular systems\cite{butcher,boyd,butcher64} where 
the position operator ${\bf r}$ is well-defined in real space.
However, for periodic systems, the usual definition of ${\bf r}$ is no longer 
valid over all space. Instead the ``saw-liked'' position operator must be 
introduced
to maintain the periodic property of the system\cite{peeters93,vanderbilt}.
If periodic boundary conditions are applied to a physical system, 
the average electronic position could be anywhere for delocalized 
states\cite{peeters93}. This is clearly not the case for
most molecular systems with only bound states. 
For periodic systems, the position operator ${\bf r}$ is conveniently
defined in momentum space\cite{blount}:
\begin{eqnarray}
{\bf r}_{n {\bf k}, n' {\bf k'}}= i {\bf \nabla_k} \zeta_{n,n'}
({\bf k},{\bf k'})
+ \Omega_{n,n'}({\bf k})\delta({\bf k}-{\bf k'}),
\label{eq:r}
\end{eqnarray}
where
\begin{eqnarray}
\zeta_{n,n'}({\bf k},{\bf k'})=\int_{V} \psi_{n, {\bf k}}^*({\bf r})
\psi_{n', {\bf k'}}({\bf r}) d{\bf r}, \nonumber \\
\displaystyle \Omega_{n,n'}({\bf k})=\frac{i}{v}\int_{v}
u_{n,{\bf k}}^*({\bf r}){\bf \nabla_{k}} u_{n', {\bf k}}({\bf r}) d {\bf r},
\end{eqnarray}
$V$ is the whole system volume, $v$ is the unit cell volume, and
$\psi_{n, {\bf k}}({\bf r}) = u_{n, {\bf k}}({\bf r})e^{i{\bf k \cdot r}}$ 
is the Bloch state with $n$ and ${\bf k}$ the band index and crystal momentum, 
respectively. 
The two terms in Eq.\eqref{eq:r} correspond to the intra- and inter-band
transitions, respectively\cite{blount}.

On the other hand, the ${\bf p\cdot A}$ (current-current correlation)
instead of the ${\bf E\cdot r}$ (dipole-dipole correlation) gauge is
often used, where ${\bf p}$ is treated as static\cite{butcher,peeters93}.
Though quite successful in the linear case, the static
current-current correlation is actually not equivalent to the dipole-dipole
correlation (i.e., the ${\bf E\cdot r}$) gauge and gives the wrong results
for periodic systems\cite{mxu3,mxu4} in nonlinear optical studies.
To restore 
the equivalence 
between these two gauges, one needs to incorporate the proper gauge phase 
factor in the current-current correlation\cite{mxu3}. The importance of
the gauge phase factor in gauge equivalence is also verified in a recent
work by Rzazewski and Boyd\cite{boyd04}. For simplicity, 
we have used the ${\bf E\cdot r}$ gauge in this paper.

To demonstrate the symmetry breakdown, we focus on
two 1D single-electron periodic models - 
the Su-Shrieffer-Heeger(SSH) model\cite{ssh} and
its continuum sibling Takayama-Lin-Liu-Maki(TLM) model\cite{tlm}. These two models 
are quite successful in 
explaining optical properties of trans-polyacetylene (see, for example,
\cite{heeger88}). The advantages of 1D single-electron
periodic models are two folds. First, since $\zeta_{n,n'}({\bf k},{\bf
  k'})=\delta_{n,n'}\delta({\bf k}-{\bf k'})$ 
can be applied in Eq.\eqref{eq:r},
these two models result in 1D simple bands and are analytically 
solvable\cite{kohn59,zak85}.
Second, single-electron models also
satisfy physical conditions where the overall permutations symmetry
is supposed to be strictly held in existing 
theories\cite{butcher,boyd,butcher64}, since the medium is loss-free at all the 
relevant optical frequencies under these models.
When centro- or inversion symmetry is applied to the system, 
$\chi^{(2)}$ vanishes and $\chi^{(3)}$ becomes the first non-zero nonlinear 
susceptibility.
Hence, we will concentrate our discussion on $\chi^{(3)}$. 

The Hamiltonian of the SSH model with the rigid-lattice approximation
(i.e., the dimerized H\"{u}ckel model) is described as follows\cite{ssh}: 
\begin{eqnarray}
H_{SSH}^0=-\sum_{l,s} \left[ t_0+(-1)^l \frac{\Delta}{2} \right]
(\hat{C}_{l+1,s}^{\dag}\hat{C}_{l,s}^{}+\hat{C}_{l,s}^{\dag}\hat{C}_{l+1,s})^{},
\label{eq:Hssh}
\end{eqnarray}
where $t_0$ is the transfer integral between the nearest-neighbor sites,
$\Delta$ is the gap parameter and $\hat{C}_{l,s}^{\dag}(\hat{C}_{l,s})$
creates(annihilates) a $\pi$ electron at site $l$ with spin $s$.
Following the same procedure described in previous
work\cite{mxu1,mxu2,mxu3,mxu4,jiang1}, we consider the momentum space
representation of the Hamiltonian
given by Eq.\eqref{eq:Hssh}. With the aid of the spinor description
$\hat{\psi}_{k,s}^{\dag}(t)$=$(\hat{a}^{\dag c}_{k,s}(t)$,
$\hat{a}^{\dag v}_{k,s}(t))$, where $\hat{a}^{\dag c}_{k,s}(t)$ and
$\hat{a}^{\dag v}_{k,s}(t)$ are the excitations of electrons in the conduction
band and the valence band with momentum $k$ and spin $s$, we obtain the
following formula:
\begin{eqnarray}
\hat{H}_{SSH}(k,t)&=&\hat{H}_{SSH}^0+\hat{H}_{\bf E \cdot r} \nonumber \\ 
&=&\sum_{-\frac{\pi}{2a}\le k\le\frac{\pi}{2a},s} \varepsilon(k)
\hat{\psi}_{k,s}^{\dag}(t) \sigma_3 \hat{\psi}_{k,s}(t)
- \hat{D} \cdot E_0 e^{i\omega t},
\label{eq:Hsshk}
\end{eqnarray}
where
\begin{eqnarray}
\hat{D}= e \sum_{-\frac{\pi}{2a}\le k\le\frac{\pi}{2a},s}
(\beta(k)\, \hat{\psi}_{k,s}^{\dag} \sigma_2\hat{\psi}_{k,s}
 +i \frac{\partial}{\partial k} \, \hat{\psi}_{k,s}^{\dag}\hat{\psi}_{k,s}),
\label{eq:D}
\end{eqnarray}
$\varepsilon (k)= \sqrt{\left[ 2 t_0 cos(ka) \right]^2+\left[ \Delta sin(ka)
\right]^2}$,
$\sigma_i$ ($i=1,2,3$) are Pauli matrices, and  
$\beta(k)=-\Delta t_0 a / \varepsilon^2(k)$. $\beta(k)$
is the coefficient related to the interband transition between the conduction
and valence bands in a unit cell of length $2a$, and the second term in
Eq.\eqref{eq:D} is related to the intraband 
transition\cite{mxu1,mxu2,mxu3,mxu4,jiang1}.

According to field theory and the Wick theorem\cite{mahan1}, the general 
four-wave-mixing(FWM) can be expressed as\cite{mxu1,mxu2,mxu3,mxu4,jiang1}:
\begin{eqnarray}
\chi_{SSH}^{(3)}(-\Omega;\omega_1,\omega_2,\omega_3)&=& 
\frac{2 e^4 n_0}{\hbar^3}
\frac{1}{3!L}\sum_{k,{\mathcal{P}}(\omega_1,\omega_2,\omega_3)}\int
\frac{id\omega}{2\pi}Tr\Biggl\{ (\beta(k)\sigma_2
+i\frac{\partial}{\partial k}) G(k,\omega) \nonumber \\
& &(\beta(k)\sigma_2+i\frac{\partial}{\partial k}) G(k,\omega-\omega_1)
(\beta(k)\sigma_2+i\frac{\partial}{\partial k}) G(k,\omega-\omega_1-\omega_2)
\nonumber \\
& &(\beta(k)\sigma_2+i\frac{\partial}{\partial k}) G(k,\omega-\omega_1-\omega_2
-\omega_3)
\Biggr\},
\label{eq:fwm1}
\end{eqnarray}
where $\Omega=\omega_1+\omega_2+\omega_3$, $L$ is the chain length, $n_0$ is the
number of chains per unit cross area, and
${\mathcal{P}}(\omega_1,\omega_2,\omega_3)$ represents all six permutations over
$\omega_1$, $\omega_2$ and $\omega_3$. The polymer chains are assumed to
be oriented. Green's function $G(k,\omega)$ is defined as
follows\cite{mxu1,mxu2,mxu3}:
\begin{eqnarray}
\displaystyle G(k,\omega)=
\frac{\omega+\omega_k\sigma_3}{\omega^2-\omega_k^2+i\epsilon},
\label{eq:green}
\end{eqnarray}
with $\omega_k \equiv \varepsilon(k) / \hbar \text{ and } \epsilon \equiv 0^+$.

The analytical expression for the general four-wave-mixing can be
found in \cite{jiang1}:
\begin{equation}\label{eq:FWMSSH}
\begin{aligned}
\chi^{(3)}_{SSH}(-\Omega;\omega_1,\omega_2,\omega_3)
&=\chi_{0}^{(3)} \frac{15}{1024}
\sum_{{\mathcal{P}}(z_1,z_2,z_3)}
 \int_{1}^{1/\delta}\frac{xdx}
{\sqrt{(1-\delta^{2}x^2)(x^2-1)}} \\
&\left\{-\frac{(2x-z_{1}-z_{3})}
{x^{8}(x-z_{1})(x+z_{2})(x-z_{3})(x-z_{1}-z_{2}-z_{3})}\right.\\
&-\frac{(2x+z_{1}+z_{3})}
{x^{8}(x+z_{1})(x-z_{2})(x+z_{3})(x+z_{1}+z_{2}+z_{3})}\\
&+\frac{4(1-\delta^{2}x^2)(x^2-1)}{x^8(x-z_{1}-z_{2})}
\frac{(3x-2z_{1})(3x-2(z_{1}+z_{2}+z_{3}))}
{(x-z_{1})^{2}(x-z_{1}-z_{2}-z_{3})^2}\\
&+\left.\frac{4(1-\delta^{2}x^2)(x^2-1)}{x^8(x+z_{1}+z_{2})}
\frac{(3x+2z_{1})(3x+2(z_{1}+z_{2}+z_{3}))}
{(x+z_{1})^{2}(x+z_{1}+z_{2}+z_{3})^2}\right\},
\end{aligned}\end{equation}
where $\displaystyle \chi_0^{(3)} \equiv \frac{8}{45}\frac{e^4 n_0}{\pi}
\frac{(2 t_0 a)^3}{\Delta^6}$, 
$\displaystyle z_i\equiv\frac{\hbar\omega_i}{2\Delta} (i=1,2,3)$, 
$\displaystyle \delta=\frac{\Delta}{2t_0}$, and
${\mathcal{P}}(z_1,z_2,z_3)$ represents all six permutations over
$z_1$, $z_2$ and $z_3$ (Please note that the summation over all six
permutations was missing in the original Eq. (3.3) in
Ref. \cite{jiang1} due to a typo.).

To discuss the breaking of the overall permutation symmetry, we may
simply consider the difference
\begin{equation}
\label{eq:diff}
\delta \chi^{(3)}=\chi_{SSH}^{(3)}(-\Omega;\omega_1,\omega_2,\omega_3)
-\chi_{SSH}^{(3)}(\omega_1;-\Omega,\omega_2,\omega_3).
\end{equation}
Due to the trivial permutation of the spatial indices,
the spatial indices $\mu\alpha_1\cdots\alpha_n$
can be dropped in the expression of $\chi^{(n)}$ in the 1D case. Thus, the 
Kleinman 
symmetry is exactly the same as the overall permutation symmetry 
for the 1D case, though it is generally different 
 from the overall permutation symmetry for higher dimensional cases.
Obviously, $\delta\chi^{(3)}$ in Eq.\eqref{eq:diff} should be identically 
{\it zero} if the overall permutation symmetry were preserved.
However, it is easy to see from \eqref{eq:FWMSSH} 
that $\chi_{SSH}^{(3)}(-\Omega;\omega_1,\omega_2,\omega_3)$
is symmetric in $\omega_1$, $\omega_2$, and $\omega_3$, while 
$\chi_{SSH}^{(3)}(\omega_1;-\Omega,\omega_2,\omega_3)$ is only symmetric in
$\omega_2$ and $\omega_3$. Hence, $\delta \chi^{(3)}$ can not be zero everywhere.
Indeed, replacing $\omega_1$ by $-(\omega_1+\omega_2+\omega_3)$ in 
\eqref{eq:FWMSSH}, substituting the resulting expression into
\eqref{eq:diff}, and then simplifying the final expression, we obtain
\begin{equation}
\label{eq:broken}
\begin{aligned}
\delta \chi^{(3)}
&=\chi^{(3)}_{SSH}(-(\omega_1+\omega_2+\omega_3);\omega_1,\omega_2,\omega_3)
-\chi^{(3)}_{SSH}(\omega_1; -(\omega_1+\omega_2+\omega_3),\omega_2,\omega_3)\\
&=\chi_{0}^{(3)} \frac{15}{256}\sum_{{\mathcal{P}}(z_2,z_3)}
 \int_{1}^{1/\delta}\frac{\sqrt{(1-\delta^{2}x^2)(x^2-1)}dx}{x^7}\cdot \\
&\left\{
\frac{(3x+2z_2)(3x+2(z_1+z_2+z_3))}{(x+z_2)^2(x+z_1+z_2+z_3)^2}
\left(\frac{1}{x+z_1+z_2}+\frac{1}{x+z_2+z_3}\right)\right.\\
&\left.+
\frac{(3x-2z_2)(3x-2(z_1+z_2+z_3))}{(x-z_2)^2(x-z_1-z_2-z_3)^2}
\left(\frac{1}{x-z_1-z_2}+\frac{1}{x-z_2-z_3}\right)\right.\\
&-\frac{(3x+2z_1)(3x-2z_2)}{(x+z_1)^2(x-z_2)^2}\left(\frac{1}{x+z_1+z_3}
+\frac{1}{x-z_2-z_3}\right)\\
&\left.-\frac{(3x-2z_1)(3x+2z_2)}{(x-z_1)^2(x+z_2)^2}\left(\frac{1}{x-z_1-z_3}
+\frac{1}{x+z_2+z_3}\right)\right\},
\\
\end{aligned}
\end{equation}
where the summation is over the two permutations of $z_2$ and $z_3$. 
Eq. \eqref{eq:broken} provides a general expression for the deviation of the 
overall permutation symmetry under the SSH or 1D dimerized H\"{u}ckel model.

\begin{rem}
A closer examination shows that the first two terms in
\eqref{eq:FWMSSH} come from the interband transition, and
the last two terms in \eqref{eq:FWMSSH} come from the intraband
transition. The deviation of the overall permutation symmetry 
(Eq. \eqref{eq:broken}) is due to the intraband transition alone.
\end{rem}

To demonstrate the large deviation of the overall permutation symmetry 
quantitatively, we may consider two special cases under the TLM model 
(the continuum limit of the SSH model):
$\chi_{TLM}^{(3)}(-3\omega;\omega,\omega,\omega)$ and
$\chi_{TLM}^{(3)}(\omega;\omega,\omega,-3\omega)$.
Our analysis shows that 
\begin{equation}
\begin{aligned}
\chi_{TLM}^{(3)}(-3\omega;\omega,\omega,\omega)
&=\chi_0^{(3)} \frac{45}{128} \, \Biggl\{ -\frac{14}{3 z^8}-\frac{4}{15 z^4}
+\frac{(37-24 z^2)}{8 z^8} f(z)+\frac{(1-8 z^2)}{24 z^8} f(3z) \Biggr\}\\
&= \chi_{0}^{(3)} (\frac{5}{28}+\frac{80}{11}z^2+\frac{98580}{1001}z^4 +O(z^6))
\ \ \ (z\rightarrow 0),
\label{eq:tlmthg}
\end{aligned}
\end{equation}
and
\begin{equation}
\begin{aligned}
\chi_{TLM}^{(3)}(\omega;\omega,\omega,-3\omega)
&=\chi_0^{(3)}\frac{5}{1024z^8}\Biggl\{\frac{5}{3}(40z^2-61)f(z)+
\frac{16}{3}(4z^2-1)f(2z)\Biggr. \\
&\Biggl.-\frac{1}{243}(1944z^2-241)f(3z)
+\frac{32}{243}(27z^4-30z^2+805)\Biggr\} \\
&=\chi_{0}^{(3)}(\frac{5}{28}+\frac{80}{33}z^2+\frac{28500}{1001}z^4 +O(z^6))
\ \ \ (z\rightarrow 0).
\label{eq:kleinman}
\end{aligned}
\end{equation}
where $z=\hbar \omega/2\Delta$, and the function $f$ is defined by the formula
\begin{equation}
f(z) \equiv \left \{
\begin{array}{lr}
\displaystyle \frac{\arcsin (z)}{z \sqrt{1-z^2}},  &(z^2<1)\\
\displaystyle -\frac{\cosh^{-1} (z)}{z\sqrt{z^2-1}}+\frac{i\pi}
{2 z\sqrt{z^2-1}}, &(z^2>1).
\end{array}
\right.
\label{eq:fz}
\end{equation}
For polyacetylene, using typical parameters of $t_0=2.5 eV$, $\Delta=0.9 eV$, 
$n_0=3.2 \times 10^{14} cm^{-2}$ and $a=1.22 \AA$ results in
$\chi_0^{(3)} \approx 1.0 \times 10^{-10}$ 
esu\cite{mxu1,mxu2,mxu3,mxu4,jiang1}. 
In Fig.\ref{gr:fig_overall} 
we have plotted
$\chi_{TLM}^{(3)}(-3\omega;\omega,\omega,\omega)$, 
$\chi_{TLM}^{(3)}(\omega;\omega,\omega,-3\omega)$ and the difference between
the above two $\chi^{(3)}$ in the off-resonant region.
The graph shows that there is about a 40\% difference for $z=1/6$ (about
$0.3eV$ or $4.14\mu m$) between these two quantities. 
Subtracting the two asymptotic expressions in Eq. \eqref{eq:tlmthg} and 
\eqref{eq:kleinman} reveals that the
difference between $\chi_{TLM}^{(3)}(-3\omega;\omega,\omega,\omega)$ and 
$\chi_{TLM}^{(3)}(\omega;\omega,\omega,-3\omega)$ in the off-resonant
region satifies the following relationship:
\begin{equation}
\delta\chi^{(3)}(\omega) \propto \displaystyle \frac{e^4n_0t_0^3a^3
\hbar^2\omega^2}{\Delta^8}.
\label{eq:levine}
\end{equation}

We have also computed nonlinear susceptibilities without the $\nabla_k$ or 
$\partial_k$ terms (corresponding to the intraband transitions) in
Eq.\eqref{eq:D}\cite{jiang1}. The results preserve the overall permutation 
symmetry.
Excluding the gradient terms is a crude approximation for molecular systems 
where the polarization current is not present\cite{kittel}.
Therefore in periodic systems, it is the gradient term that breaks the overall 
permutation symmetry which remains valid in molecular systems
\cite{butcher,boyd,butcher64}. Obviously, this is closely related to the fact 
that the position operator for periodic systems is entirely different
from that for molecular systems.

\begin{rem}
Recently, deviations from Kleinman symmetry in the low-frequency off-resonant
regions have been observed in many nonlinear optical 
experiments\cite{crane76,singh72,chemla72,lopez98,tsutsumi98,tsutsumi99,%
wortmann92,rojo98,shelton85,shelton88}. Upon
a careful examination of these experiments, we observed that: 
(i). The deviation of the Kleinman symmetry increases with decreasing band gap
and is proportional to $\omega^2$ for crystals \cite{crane76};
(ii). the deviation of the Kleinman symmetry in
delocalized states such as aromatic 
molecules\cite{wortmann92,rojo98} and some 
polymers\cite{lopez98,tsutsumi98,tsutsumi99} or 
crystals\cite{crane76,singh72,chemla72} is usually much larger than 
that in localized states such as molecular systems such as $O_2$, $N_2$, 
etc\cite{shelton85,shelton88} (20$\sim$50\% versus $\le$8\%).
Eq.\eqref{eq:levine} captures the dependence of $\omega^2$ and the bad gap
observed in (i). Furthermore, since
the overall permutation symmetry is the basis of the Kleinman symmetry, our 
computation
also provides a natural explanation about (ii), as we have shown
that the overall permutation symmetry is broken in periodic systems. 
Finally, the vanishing $\chi^{(2)}$ 
under the SSH or TLM model shows that some symmetries such as centro-symmetry
can suppress the deviation from Kleinman symmetry even for periodic
systems. This may explain why Kleinman symmetry is still preserved
in some $\chi^{(2)}$ experiments of crystals\cite{parsens71}.

Kleinman symmetry (or Kleinman conjecture) is considered
valid only when $\omega=0$\cite{butcher,boyd}. 
Previous theoretical
calculations\cite{garito79,karna92,rashkeev98} have explicitly shown
the Kleinman symmetry breakdown when $\omega\ne0$ for higher dimensional
cases. Nevertheless, it
is still generally treated as a valid approximate symmetry for all physical
systems in low-frequency off-resonant regions.
However, many experiments show that the range of Kleinman symmetry is
substantially more restrictive than what is widely assumed in theory, especially
for periodic systems\cite{simpson04}. To explain the large Kleinman
symmetry deviation in periodic systems
\cite{crane76,singh72,chemla72,lopez98,tsutsumi98,tsutsumi99,wortmann92,%
rojo98,shelton85,shelton88},
various models have been presented. For example, Levine's model that
predicts the second order polarizability tensor $d^F/d^A\propto\omega^2$
where superscripts F and A mean forbidden and allowed 
respectively\cite{levine73},
dipole contributions from two perpendicular directions\cite{wortmann92},
harmonic frequency $2\omega$ strongly resonant with the Q band\cite{rojo98},
mutual exclusion properties between Kramers-Kronig dispersion relations
and Kleinman symmetry\cite{simpson04}, etc. Without so many input parameters,
the overall permutation symmetry breakdown in periodic systems 
provides a much more straightforward and general explanation of the above 
experiments. 
\end{rem}

In conclusion, the overall permutation symmetry for nonlinear 
susceptibilities is, albeit preserved in bound-state molecular systems, no 
longer generally held in periodic systems with
delocalized states. Therefore, it leads to large deviations of Kleinman 
symmetry. The deviation $\delta\chi^{(3)}$ is proportional to
$(e^4n_0t_0^3a^3\hbar^2\omega^2)/\Delta^8$ in the off-resonant regions. 
Theoretically, non-interacting centro-symmetric 1D periodic structures 
such as single crystals of conjugated molecules are ideal
materials for performing off-resonant $\chi^{(3)}$
experiments to directly test the overall permutation symmetry breakdown.
Practically, electron-electron correlation and other interactions are 
very important in 1D conjugated systems\cite{heeger88}. These
interactions will certainly change the magnitude of the deviation.
However, as long as those
interactions are not strong enough to destroy the band structures where the
intraband transitions remain valid, the overall permutation symmetry 
breakdown can still be observed.

\acknowledgements The authors would like to thank Dr. Xing Wei for very 
helpful discussions.
\begin{figure}[tbp]
\vskip -10pt
\begin{center}
\epsfxsize=12cm \epsfbox{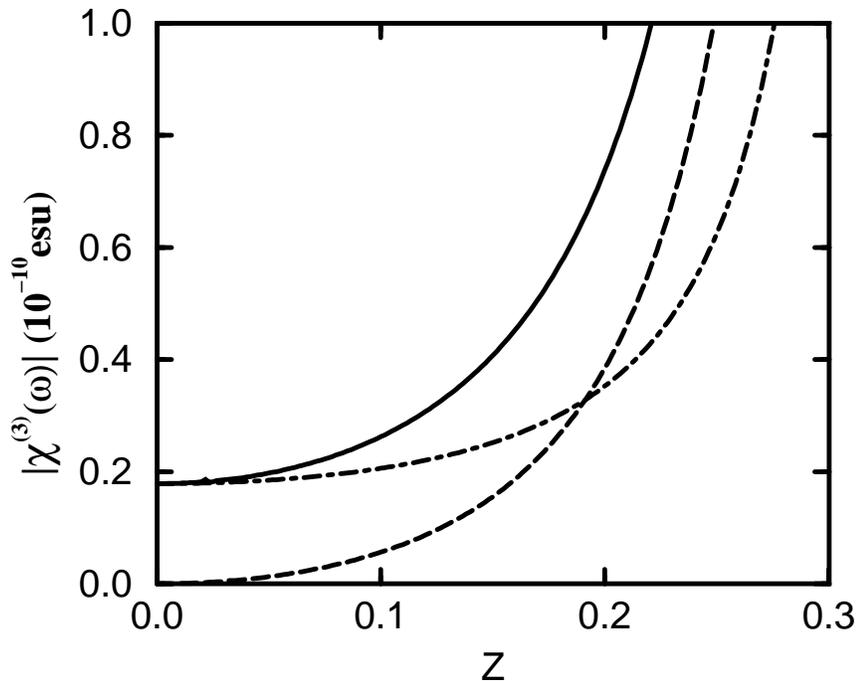}
\end{center}
\vskip 0pt
\caption{Hyperpolarizabilities under the TLM model in the off-resonant region:
$\chi^{(3)}_{TLM}(-3\omega;\omega,\omega,\omega)$ (solid line),
$\chi^{(3)}_{TLM}(\omega;\omega,\omega,-3\omega)$ (dot dashed line), and 
their difference (long dashed line); the horizontal axis is defined by
$Z\equiv\hbar\omega/2\Delta$.}
\label{gr:fig_overall}
\end{figure}

\bibliography{}
}
\end{document}